\documentclass[aps,pra]{revtex4}
\usepackage{amsmath}
\usepackage{amssymb}
\usepackage{graphics}
\usepackage{graphicx}
\usepackage{comment}
\usepackage{lmodern}
\usepackage{subfig}

\newcommand\ket[1]{\left|#1\right>}
\newcommand\bra[1]{\left<#1\right|}

\newcommand\expect[1]{\left<#1\right>}

\newcommand\Schr{Schr\"odinger\,}

\newcommand\dg{^\dagger}

\newcommand\Tr{\mathrm{Tr}}

\newcommand\ro{\hat\rho}
\newcommand\Ho{\hat H}
\newcommand\xo{\hat x}
\newcommand\po{\hat p}
\newcommand\si{\sigma}
\newcommand\Psit{\widetilde\Psi}
\newcommand\rot{\hat{\widetilde\rho}}

\newcommand\Mean{\mathbf{M}}

\begin{document}
\title{Notes on oldest jump unravelling of spatial decoherence master equation}  
\author{G\'abor Homa}
\email{ggg.maxwell1@gmail.com}
\affiliation{
   Department of Physics of Complex Systems,\\
 ELTE E\"otv\"os Lor\'and University,\\
   P\'azm\'any P\'eter s\'et\'any 1/A, H-1117 Budapest
   }
	\author{Lajos Di\'osi}
	\email{diosi.lajos@wigner.mta.hu}
	\affiliation{
	Wigner Research Centre for Physics,\\
	High Energy Physics Department\\
H-1525 Budapest 114., P.O.B. 49, Hungary}
\date{\today}

\begin{abstract}
Solution of free particle quantum master equation with spatial decoherence can be unravelled into
stochastic quantum trajectories in many ways. The first example, published in 1985, proposed a
piecewise deterministic jump process for the wave function. While alternative unravellings, diffusive
ones in particular, proved to be tractable analytically, the jump process 1985, also called orthojump, 
allows for few analytic results, it needs numeric methods as well. 
Here we prove that, similarly to diffusive unravellings, it is localizing the quantum state. 
\end{abstract}
\maketitle
\emph{Introduction.}
A single \Schr particle becomes a simple open quantum system if the particle is interacting with a thermal reservoir.
Its dynamics is given by a master equation which can take the following simple form valid typically at high temperatures:
\begin{equation}\label{ME}
\frac{d\ro}{dt}=-\frac{i}{\hbar}[\Ho,\ro]-\frac{D}{\hbar^2}[\xo,[\xo,\ro]],
\end{equation}
where $\Ho=(\po^2/2m)$ is the particle's Hamiltonian, $\xo,\po$ are its coordinate and momentum resp., and $D$ is the diffusion constant.
Joos and Zeh suggested this equation as the simplest model of spatial decoherence \cite{JooZeh85} while at the time similar
single particle master equations were known from various fields, cf., e.g., \cite{Dek81,CalLeg83} . The Wigner function of $\ro$
satisfies the classical Fokker-Planck equation in the high-temperature limit:
\begin{equation}\label{FP}
\frac{d\rho(x,p)}{dt}=-\frac{p}{m}\partial_x\rho(x,p)-D\partial_p^2\rho(x,p).
\end{equation}
This elucidates the importance of the master equation \eqref{ME} as the quantized version of diffusion.
Accordingly, $D$ is the coefficient of spatial decoherence as well as of momentum diffusion: 
the two effects are alternative interpretations of the non-Hamiltonian mechanism in the master equation.
It is well-known that the classical diffusion \eqref{FP} can be equivalently described by random trajectories $(x_t,p_t)$
in phase space. The same concept applies to the master equation \eqref{ME} as well. The stochastic \emph{quantum
trajectories} are featured by state vectors $\Psi_t$ evolving by a stochastic process such that the stochastic mean
\begin{equation}\label{unr}
\Mean\Psi_t\Psi_t\dg=\ro_t
\end{equation}
satisfies the master equation \eqref{ME}. Then the quantum trajectories $\Psi_t$ are said to \emph{unravel}
the master equation. 

The unravelling is never unique, one can choose diffusive unravellings, jump unravellings, or even their combinations.
The earliest unravelling was the orthojump process \cite{Dio85}.
It turned out subsequently that any master equation possesses a standard jump and a standard diffusive unravelling \cite{Dio86a}. All possible
diffusive unravellings can be parametrized uniquely \cite{RigMotOma97,WisDio01}, each of them corresponds to a given structure of time-continuous
monitoring the  system in question \cite{WisDio01}. Similar classification is still missing
for jump unravellings.   

While quantum trajectories became
instrumental soon for quantum optics \cite{DalCasMol89,Car93,MilBresWis95}, their invention happened earlier in studies of foundations.
In the nineteen-eighties, diffusive quantum trajectories were invented by Gisin  to model quantum state collapse in a discrete
system \cite{Gis84}. One of the present authors constructed jump \cite{Dio85} and diffusive \cite{Dio88a} unravellings of the master equation \eqref{ME} 
for his gravity-related spontaneous state collapse theories \cite{Dio86b} and \cite{Dio89}, respectively. (On three decades of various spontaneous collapse theories, all based on 
unravellings, see the recent review by Bassi et al \cite{Basetal13}). 

Analytic proof was found for the wave function localization in
diffusive quantum trajectories \cite{Dio88b}. The wave function is approaching a steady localized shape for long times, as we recapitulate it below. 
Localization in the specific jump unravellings \cite{Dio85} has, however, never been
studied. The problem is more complicated than the diffusive case because jumps will never allow for a steady shape.
An analytic proof of localization has not yet been found, we shall rely on numeric (Monte-Carlo) simulations.
Jump quantum trajectories of spatial decoherence were carefully studied by Gisin and Rigo \cite{RigGis96}, 
and in a sequence of works by Hornberger and co-workers \cite{BusHor0910,LucHor13,SorHor15} for modifications of the master equation \eqref{ME} which
included friction.  Due to friction, quantum trajectories did reach a localized steady shape, calculable analytically.
The effect and proof was bound to the presence of friction. Localization in the frictionless case \eqref{ME} has remained to
be studied in the present work.

We are going to study localization of quantum trajectories in both position and momentum. Consider the
unitary transformation of a state $\Psi$ to its  centre-of-mass frame:
\begin{equation}\label{Psicom}
\Psit=\exp\left(i\expect{\xo}\po-i\expect{\po}\xo\right)\Psi_t,
\end{equation}
where the centre-of-mass state satisfies $\langle\Psit\vert\xo\vert\Psit\rangle=0$ and $\langle\Psit\vert\po\vert\Psit\rangle=0$ 
by construction. Now we can define the centre-of-mass density matrix as follows:
\begin{equation}\label{rhocom}
\Mean\Psit_t\Psit_t\dg=\rot_t.
\end{equation}
This matrix is non-negative and of unit trace, like common density matrices. Its evolution, however, is
non-linear, completely different from the master equation \eqref{ME} of the common density matrix $\ro_t$. We use $\rot_t$ to characterize
average localization of quantum trajectories $\Psi_t$ around their individual centre-of-mass $\expect{\xo}_t,\expect{\po}_t$. 
We can define centre-of-mass (squared) spreads by $(\Delta\widetilde x)^2=\Tr(\xo^2\rot)$ and by  $(\Delta\widetilde p)^2=\Tr(\po^2\rot)$.

\emph{Diffusive unravelling.} 
Following \cite{Dio88a,Dio88b}, consider the stochastic \Schr equation \cite{Dio88b}:
\begin{equation}\label{SSEd}
d\Psi= -\frac{i}{\hbar}\Ho\Psi dt -\frac{D}{\hbar^2}(\xo-\expect{\xo})^2\Psi+\frac{\sqrt{2D}}{\hbar}(\xo-\expect{\xo})\Psi dW,
\end{equation}
where $dW$ is the Ito-differential of the Wiener stochastic process, satisfying $\Mean dW=0$, $(dW)^2=dt$.
The solutions satisfy the condition \eqref{unr} of unravelling. 
For long, the centre-of-mass solutions converge to the following complex Gaussian wave packet: 
\begin{equation}\label{Psistac}
\Psit_\infty(x)=\frac{1}{(2\pi\si_\infty^2)^{1/4}}\exp\left(-(1-i)\frac{x^2}{4\si_\infty^2}\right) 
\end{equation}li
of squared width
\begin{equation}
\si_\infty^2=\sqrt{\frac{\hbar^3}{2Dm}}.
\end{equation}
According to \eqref{Psistac}, the centre-of-mass density matrix \eqref{rhocom} turns out to converge to a pure state:
\begin{equation}\label{rhocom}
\rot_\infty=\Psit_\infty\Psit_\infty\dg.
\end{equation}
The coordinate and momentum spreads are given by
\begin{equation}\label{DxDpdiff}
(\Delta\widetilde x)^2=\sigma_\infty^2,~~~~(\Delta\widetilde p)^2=\frac{\hbar^2}{2\sigma_\infty^2}.
\end{equation}
The centre-of-mass of $\rot_\infty$ keeps to perform the following diffusive motion:
\begin{equation}\label{comstac}
d\expect{\xo}=\frac{1}{m}\expect{\po}dt+\sqrt{\frac{2\hbar}{m}}dW,~~~~~~~~d\expect{\po}=\sqrt{2D}dW.
\end{equation}
Observe that the diffusion of the momentum is the classical one. On the contrary, the diffusion of the coordinate cannot
happen classically, it is purely quantum. Asymptotic localization is thus the analytically calculable feature of the diffusive 
quantum trajectories of the simple spatial decoherence master equation \eqref{MeE}.

\emph{Orthojump unravelling.} 
For the sake of comparison with the diffusive unravelling, let us cast the jump unravelling of \cite{Dio85} 
into the  form of a stochastic  \Schr equation:  
\begin{equation}\label{SSEj}
d\Psi=-\frac{i}{\hbar}\Ho\Psi dt-\frac{D}{\hbar^2}[(\xo-\expect{\xo})^2-\si^2]\Psi dt+\left(\frac{x-\expect{\xo}}{\si}-1\right)\Psi dN,
\end{equation}
where $\si^2=\langle(\xo-\expect{\xo})^2\rangle$. $dN$ stands for the Ito-differential of a Poisson process, satisfying $\Mean dN=2D\si^2 dt$, $(dN)^2=dN$.
This equation corresponds to a piece-wise deterministic evolution of $\Psi_t$, interrupted by jumps at random times.
In elementary terms, the mechanism is the following.
Consider the deterministic non-linear \Schr equation
\begin{equation}\label{NLSE}
\frac{d\Phi}{dt}=-\frac{i}{\hbar}\Ho\Phi-\frac{D}{\hbar^2}[(\xo-\expect{\xo})^2]\Phi.
\end{equation}
[Note that this equation coincides with the deterministic part of the diffusive stochastic \Schr equation \eqref{SSEd} and they share 
$\Psit_\infty$ \eqref{Psistac} as  (normalized) steady-shape centre-of-mass solution.]
Solve this non-linear \Schr equation for the initial condition $\Phi_0=\Psi_0$ and 
define the physical quantum state by $\Psi_t=\Phi_t/\Vert\Psi_t\Vert$. Note
that the norm of $\Phi$ is strictly decreasing:
\begin{equation}\label{norm}
\frac{d\Vert\Phi\Vert^2}{dt}=-\frac{2D}{\hbar^2}\si^2.
\end{equation}
The probability of jump-free deterministic evolution is  decreasing exactly with the norm $\Vert\Phi\Vert^2$,
i.e., the probability rate of jump is  $(2D/\hbar^2)\si^2$. When a jump occurs,  the smooth deterministic
evolution of $\Psi/\Vert\Psi\Vert$ is interrupted by the sudden change
\begin{equation}\label{orthojump}
\Phi\longrightarrow (\xo-\expect{\xo})\Phi,
\end{equation}
rendering the new state orthogonal to what it was before the jump (cf. also \cite{MilBresWis95}). 
After the jump, the deterministic evolution \eqref{NLSE} re-starts and continues until the next jump, etc.
\begin{figure}[h]
  \centering
	\resizebox{120 mm}{!}{
	\includegraphics[scale=10]{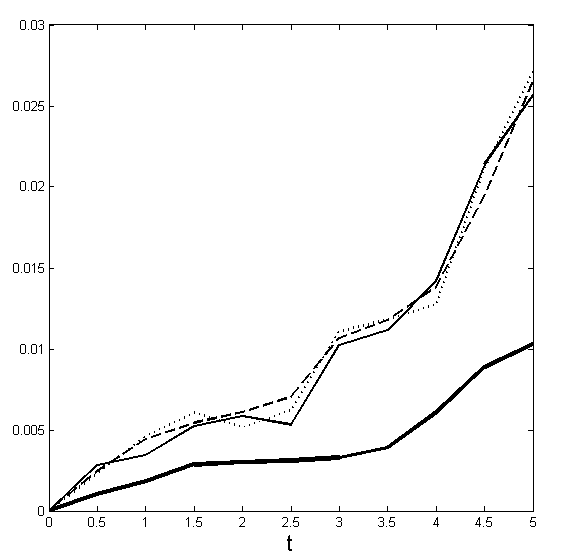}}
 \caption{
Normalized distance $\sqrt{\Tr(\ro_{MC}-\ro)^2}/\sqrt{Tr(\ro^2)}$ between MC-simulated density matrix $\ro_{MC}$ and the exact $\ro$
in the time interval $t\in(0,5)$, taken on three-times $5000$ trajectories (solid, dot, dash, resp.), and on the overall $15000$ trajectories (lower solid).}
  \label{abra:error}
\end{figure}

\emph{Numeric tests of orthojumps.} 
We have performed MC simulations of the orthojump quantum trajectories.
While analytic solutions for individual trajectories are not (yet) known, the analytic solution of the master
equation \eqref{ME} is easy \cite{UnrZur89,DioKie02}, especially for Gaussian initial states \cite{Giuetal96}. 
To check the robustness of our MC simulation, we shall compare the MC-simulated density matrix $\ro_{MC}$ 
to the analytic solution $\ro$ of the master equation \eqref{ME}. 

With suitable choice of physical units, we can always take trivial parameters $\hbar=m=D=1$ and that is what we do.
For the rest of our work, we choose the pure  state \eqref{Psistac} deliberately (just
for concreteness) as the initial pure state for \eqref{ME}. 
The analytic solution in coordinate representation reads: 
\begin{equation}\label{rhoanal}
\bra{x}\ro_t\ket{y}=\frac{1}{\sqrt{2\pi}\Sigma(t)}
\exp\left\{-\frac{1}{8\Sigma^2(t)}(x+y)^2-\frac{1+2\sqrt{2}t+2t^2+\frac{2\sqrt{2}}{3}t^3+\frac{1}{3}t^4}{8\Sigma^2(t)}(x-y)^2-i\frac{1+\sqrt{2}t+t^2}{4\Sigma^2(t)}(x^2-y^2)\right\}
\end{equation}  
where the squared spatial spread is 
\begin{equation}
\Sigma^2(t)=\frac{1}{\sqrt{2}}+t+\frac{t^2}{\sqrt{2}}+\frac{t^3}{3}.
\end{equation}
With the same initial pure state, we MC-generated $3\times5000$ quantum trajectories $\{\Psi_t^{(n)}; n=1,2,\dots,15000\}$ and determined $\ro_{MC,t}$ numerically:
\begin{equation}
\bra{x}\ro_{MC,t}\ket{y}=\frac{1}{\sum_n 1}\sum_n\Psi_t^{(n)}(x)\Psi_t^{(n)*}(y).
\end{equation}
Displaying its normalised distance 
\begin{equation}
\frac{\left(\Tr(\ro-\ro_{MC})^2\right)^{1/2}}{(\Tr\ro^2)^{1/2}}
\end{equation}
from the analytic $\ro_t$ \eqref{rhoanal} in  the range
$t\in(0,5)$,
on different statistics, one confirms the stability and precision of simulation on 15 000 trajectories (Fig. 1).
For qualitative comparison, Fig. 2 shows the MC-simulated Wigner function
and the exact one at $t=5$.
\begin{figure}[h]
  \centering
	\includegraphics[width=1\textwidth]{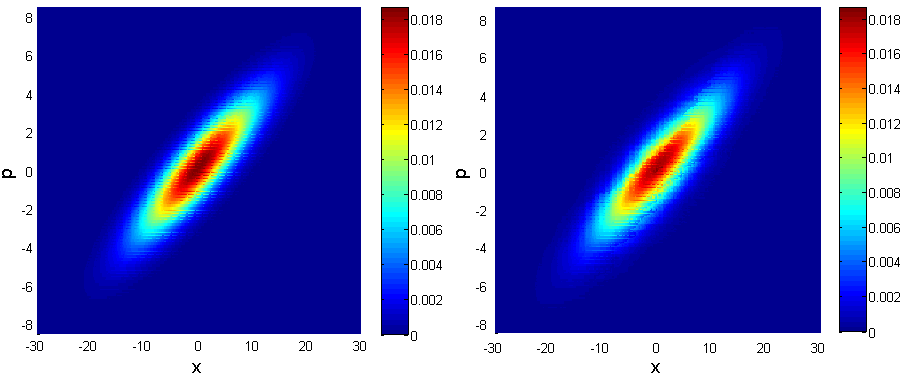}
 \caption{
Wigner function solving the master equation \eqref{ME} in units $\hbar=m=D=1$ 
at $t=5$ with initial state \eqref{Psistac}: analytic solution (left), MC solution on $15000$ trajectories (right).}
  \label{abra:error}
\end{figure}
These checks confirm that $15 000$ trajectories will suffice to
test the basic feature of interest: localization by orthojump unravelling.

We determined the centre-of-mass density matrix 
\begin{equation}\label{}
\bra{x}\rot_{CM,t}\ket{y}=\frac{1}{\sum_n 1}\sum_{n}\Psit_t^{(n)}(x)\Psit_t^{(n)*}(y)
\end{equation}
on three increasing statistics. Our main results are shown in Fig. 3, where
the time-evolution of spatial and momentum spreads 
$\Delta\widetilde x,\Delta\widetilde p$
are displayed for $t\in(0,5)$. Initial values are known analytically:
$\Delta\widetilde x_0=\Delta\widetilde p_0=1/2^{1/4}\approx0.84$. For times longer than the
characteristic time scale $1$ (when $\hbar=m=D=1$) of the master equation \eqref{ME}, localization
takes place asymptotically, both in coordinate and momentum. Both $\Delta\widetilde x$
and $\Delta\widetilde p$ converge to constants, their conservative estimates are 
\begin{equation}\label{DxDpjump}
\Delta\widetilde x_\infty=1.62\pm0.01,~~~~\Delta\widetilde p_\infty=1.63\pm0.01.
\end{equation} 
This is the first numeric evidence, in lack of analytic ones, for localization of orthojump
trajectories in frictionless spatial decoherence.
\begin{center}
\begin{figure}[]
   \centering
    \setcounter{subfigure}{0}
     \includegraphics[width=0.485\textwidth]{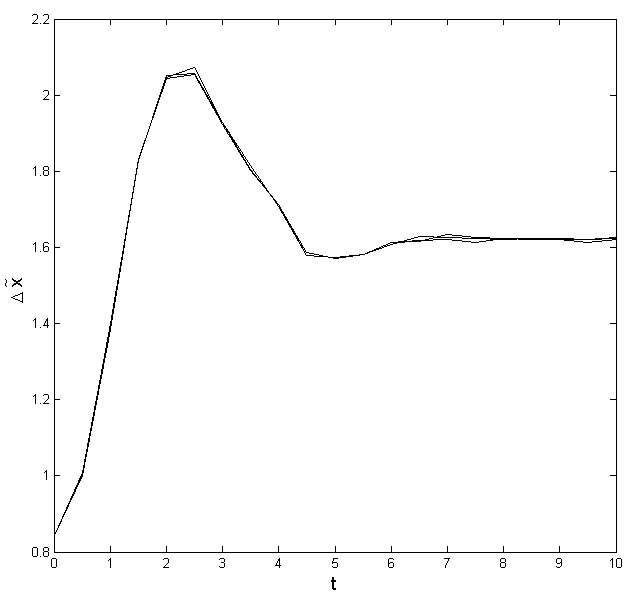}
  \quad
     \includegraphics[width=0.485\textwidth]{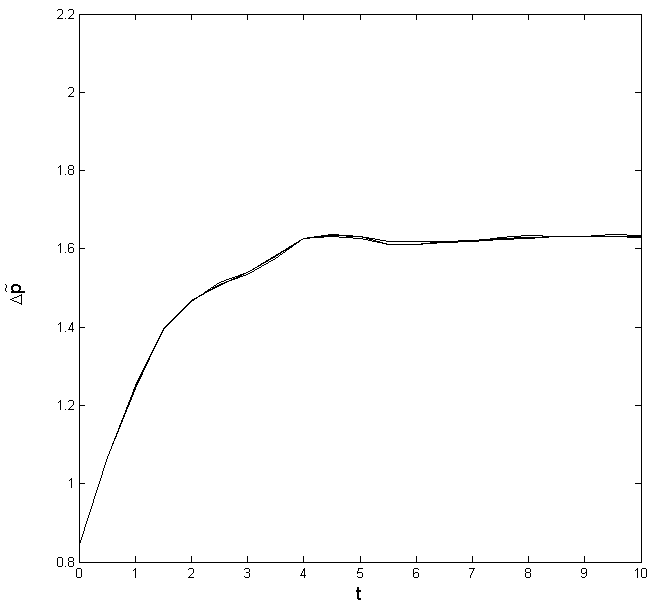}
\captionof{figure}{Centre-of-mass spreads $\Delta\widetilde x$ (a) and $\Delta\widetilde p$ (b)  in MC-simulated density matrix $\rot_{MC}$
in time interval $(0,5)$, taken on $5000$, $10000$, and $15000$ trajectories.
Values are overlapping within $0.01$.}
\end{figure}
\end{center}

\emph{Summary.}  
We have studied the localization of wave function in orthojump unravelling of the simplest and
paradigmatic spatial decoherence master equation of a free particle. Localization in diffusive
unravellings became proved analytically long ago. This time we were able to prove and
calculate localization of the orthojump unravelling --- using MC simulations. We used 15 000
MC-simulated quantum trajectories to confirm localization both in coordinate ($\Delta\widetilde x$) and 
momentum ($\Delta\widetilde p$), which we demonstrated on the centre-of-mass density matrix $\rot$. 
The obtained numeric values \eqref{DxDpjump} are about twice as large as those \eqref{DxDpdiff} in diffusive unravelling.
Such slightly looser localization may be explained  heuristically. The asymptotic centre-of-mass density matrix
$\rot_\infty$ contains randomness because the centre-of-mass wave function $\Psit_t$ never ceases to undergo
jumps, i.e., it is ``breathing'' at random times, whereas in diffusive unravelling $\Psi_t$ acquires a constant shape
for large $t$ hence $\rot_\infty$ does not contain randomness, diffusive features contribute to the centre-of-mass
motion \eqref{comstac} only. 

Our work was restricted for the demonstration of stability and localization of
the orthojump trajectories for the frictionless decoherent dynamics of a \Schr particle. Further studies should explore
more details of orthojump trajectories' rich structure. Numeric methods seem instrumeekntal. However,
similar to the diffusive case \eqref{SSEd}, a possible power of the Ito formalism \eqref{SSEj}  remains to be explored 
for analytic calculations.  

LD was supported by the Hungarian Scientific Research Fund under grant no. 103917,
and by the EU COST Actions MP1209 and CA15220.

\end{document}